\documentclass[prb,preprint,showpacs,superscriptaddress,amsfonts,amsmath,floatfix]{revtex4}

\usepackage{graphicx}
\usepackage{color}

\begin{document}

\title{Screening effect on the optical absorption in graphene and metallic monolayers}

\author{Marinko Jablan}
\email{mjablan@phy.hr}
\affiliation{Department of Physics, University of Zagreb, Bijeni\v cka c. 32, 10000 Zagreb, Croatia}

\author{Marin Solja\v ci\' c}
\email{soljacic@mit.edu}
\affiliation{Department of Physics, Massachusetts Institute of Technology, 77 Massachusetts Avenue, Cambridge MA 02139, USA}

\author{Hrvoje Buljan}
\email{hbuljan@phy.hr}
\affiliation{Department of Physics, University of Zagreb, Bijeni\v cka c. 32, 10000 Zagreb, Croatia}

\date{\today}

\begin{abstract}
Screening is one of the fundamental concepts in solid state physics. It has a
great impact on the electronic properties of graphene where huge mobilities were 
observed in spite of the large concentration of charged impurities. 
While static screening has successfully explained DC mobilities, screening properties can be significantly changed at infrared or optical frequencies. In this paper we discuss the influence of dynamical screening on the optical absorption of graphene and other 2D electron systems like metallic monolayers. This research is motivated by recent experimental results which pointed out that graphene plasmon linewidths and optical scattering rates can be much larger than scattering rates determined by DC mobilities. Specifically we discuss a process where a photon incident on a graphene plane can excite a plasmon by scattering from an impurity, or surface optical phonon of the substrate.
\end{abstract}

\maketitle

\section{Introduction}
\label{sec:intro}

In recent years there has been a lot of interest in the field of 
plasmonics which seems to be the only viable path towards realization of nanophotonics: control of light at scales substantially 
smaller than the wavelength \cite{Barnes2003}. However, plasmonic materials (most notably metals) suffer from large losses in the frequency regimes of interest, which resulted in a wide search for better materials \cite{Shalaev2010}. 
Lots of attention has recently been given to plasmonics in graphene \cite{Jablan2009, Jablan2013}, which is a single two-dimensional (2D) plane of carbon atoms arranged in a honeycomb lattice \cite{Novoselov2004,Novoselov2005a}. 
One exciting point of interest of 2D materials is that they are tunable. For example, graphene can be doped to high values of 
electron or hole concentrations by applying gate voltage \cite{Novoselov2004}, much like in field effect transistors. 
Furthermore, graphene can be be produced in very clean samples with large mobilities (demonstrated by DC transport measurements)
\cite{Novoselov2004,Novoselov2005a}. 
The DC scattering rates would imply small plasmon losses in graphene, however, it is still not clear 
how the scattering rates change with frequency, particularly in the infrared (IR) region. 
Recent nano-imaging measurements \cite{Fei2012} have demonstrated somewhat increased plasmon 
losses at IR compared to the estimate based on DC transport measurements. 
Measurements of optical transmission through graphene nano-ribbons \cite{IBM_nano_ribbon_damping} 
have demonstrated strong increase of plasmon linewidth with frequency and losses that 
are much larger than the DC estimates. However, since the ribbon width in these experiments 
is very small (10-100nm) edge scattering can significantly increase the losses. 
Nevertheless, a similar experiment \cite{Garcia_nano_rings} with graphene nano-rings has demonstrated 
plasmon linewidths that approximately agree with the DC estimate. 

Finally Electron Energy Loss Experiments (EELS) \cite{Liu2008} on graphene sheets have 
demonstrated huge plasmon linewidths that increase linearly with plasmon momentum;
however, the (DC) transport measurements were not reported so it is not clear what 
was the actual quality of the graphene films. It is also interesting to note that similar 
results \cite{Nagao2001} were obtained with EELS on the mono-atomic silver film which 
could imply a common origin of plasmon damping in these two 2D systems. 
On one hand, metallic monolayers might be even more interesting from the point of view of 
plasmonics since they have abundance of free electrons even in the intrinsic case, while 
graphene has to be doped with electrons since it is a zero band gap semiconductor. 
On the other hand, graphene has superior mechanical properties and was demonstrated in a free standing (suspended) samples while metallic monolayers have only been observed on a substrate. 

Instead of calculating plasmon linewidth, we will focus on a directly related problem 
of optical absorption, which is easier to analyze. In that respect, it was shown 
experimentally \cite{Nair2008} that suspended graphene absorbs around 2.3$\%$ 
of normal incident light in a broad range of frequencies. However, if graphene is doped with 
electrons, then Pauli principle blocks some of these transitions and there should be a 
sudden decrease of absorption below a certain threshold, which should theoretically occur at twice the Fermi energy. 
Nevertheless, optical spectroscopy experiments \cite{Basov2008} have shown that there is still lots of absorption even below this threshold. This absorption is much larger than the estimate based on DC measurements. A great deal of theoretical work addressed this problem \cite{Stauber2008,Carbotte2010,Peres2010,Vasko2012,Scharf2013}, but to our knowledge, the experimental results have quantitatively not been explained yet.

In this paper, we focus on optical absorption mediated by charged impurity scattering. 
As we have already stated, the motivation for studying this problem follows from the fact 
that typical graphene samples can have large mobilities ($\mu\approx 10000$ cm$^2/$Vs) in spite of a huge concentration of charged impurities \cite{Hwang2007_screening} ($n_i\approx 10^{12}$ cm$^{-2}$), 
which is actually comparable to the typical concentration of electrons. 
The reason one can have such a large mobility is screening \cite{Hwang2007_screening}. 
In fact, if one assumes that electrons scatter from bare charged impurities 
described with the Coulomb potential $V_q$, then the resulting mobility is almost two orders of magnitude lower than the measured value \cite{Hwang2007_screening}. 
The only way to reconcile the experiment and theory is to say that the actual scattering 
potential is screened to $V_q/\varepsilon(q)$, where $\varepsilon(q)$ is the static dielectric function. 
However, in the dynamical case, at finite frequency, screening is not so effective and 
$\varepsilon(q)$ should be replaced with the dynamic dielectric function $\varepsilon(q,\omega)$. 
This will certainly influence the single particle excitations where an incident photon excites 
an electron hole pair through impurity scattering. Moreover, at finite 
frequency one can have $\varepsilon(q,\omega)=0$ (at the plasmon dispersion) so 
there exists an additional decay channel where an incident photon excites a plasmon of the same energy, through impurity scattering. 
In other words, impurities break the translational symmetry (momentum does not need to be conserved), which allows the photon to couple directly to a plasmon mode. 
Very recently another group also calculated this process in graphene but only in the small frequency 
limit \cite{Sarma_small}. Here we give the result for the arbitrary frequency (both 
for metallic monolayers and graphene) which can be very different from the small 
frequency limit. 

More specifically, we calculate the optical absorption in the 2D electron systems 
with the randomly arranged charged impurities. First, we discuss the case of metallic 
monolayers which have a parabolic electron dispersion, and then the case of graphene with 
Dirac electron dispersion. We focus on a decay channel where the incident photon emits a plasmon 
through impurity scattering, but we also discuss a case where the incident photon emits the plasmon 
and a surface optical phonon of the substrate. For graphene on $SiO_2$ substrate, the resulting optical absorption is very small compared to the experimental results \cite{Basov2008}, and not enough to reconcile the difference between the theory \cite{Stauber2008,Carbotte2010,Peres2010,Vasko2012,Scharf2013} and the experiment\cite{Basov2008}.  
On the other hand we predict large optical absorption by plasmon emission via impurity scattering in suspended graphene. Thus we believe that these ideas can be tested in suspended graphene. Finally we note that for suspended graphene (metallic monolayers) the small frequency limit \cite{Sarma_small} gives an order of magnitude lower (larger) result than the more exact RPA calculation.

\section{Metallic monolayers}
\label{sec:metallic}

The case of the optical absorption in a bulk 3D system with parabolic electron dispersion and 
randomly arranged impurities was already studied by Hopfield \cite{Hopfield1965}. 
It is straightforward to extend his result to a 2D system and here we provide only a brief description of the calculation.  

We study a system described by the Hamiltonian $H=H_0+H_{e-e}+H_l+H_i$, where $H_0$ 
represents kinetic energy of free electrons, $H_{e-e}$ describes electron-electron 
interaction which is conveniently represented through the screening effect, $H_l$ 
describes scattering with light, and $H_i$ scattering with impurities. 
Electrons in a metallic monolayer can be described with a parabolic dispersion:
$H_0={\bf p}^2/2m^*$, where $\bf p$ is the electron momentum, and $m^*$ is effective 
mass of the electron. Next, let us introduce a monochromatic light beam of frequency 
$\omega$ which is described by the electric field  ${\bf E}(t)={\bf E}_0 e^{-i\omega t} + c.c.$. This wave is incident normally on a 2D electron gas, that is ${\bf E}(t)$ is in the plane of the gas. 
If we are only interested in a linear response with respect to this electric field, 
then interaction of electrons with light takes a particularly simple expression: 
$H_l=-i\frac{e}{m^*\omega}{\bf p}\cdot {\bf E}_0 e^{-i\omega t} +c.c.$, where 
we have introduced electron charge ($-e$). 
Further on, since momentum is a good quantum number even in an interacting electron system, 
light scattering ($H_l$) will not change the many-body eigenstates of $H_0+H_{e-e}$, but only 
the eigenvalues, see Ref. \cite{Hopfield1965}. Then, one only needs to do the perturbation 
theory in the impurity scattering. Unfortunately this trick (due to Hopfield) works only in the 
systems with parabolic electron dispersion, while in the case of Dirac electrons, like those 
found in graphene, one needs to do the perturbation theory both in the light scattering and in the impurity scattering, which is a much more tedious task.

We can write the Hamiltonian for impurity scattering as a Fourier sum over wavevectors 
$\bf q$: $H_i=\frac{1}{\Omega}\sum_{\bf q} V_i(q)e^{i{\bf q}\cdot{\bf r}}$, where $\Omega$ is the total area of our 2D system. 
By calculating the induced current to the second order in $V_i({\bf q})$ one can find the real part of conductivity \cite{Hopfield1965}:
\begin{equation}
\Re\sigma(\omega)=-\frac{e^2}{m^{*2}\omega^3}\frac{1}{\Omega}\sum_{\bf q} q_x^2  
\frac{1}{\Omega} |V_i(q)|^2 \frac{1}{V_c(q)} 
\cdot \Im\frac{1}{\varepsilon(q,\omega)}
\label{Absorp2D}
\end{equation}
Note that this quantity ($\Re\sigma(\omega)$) determines the optical absorption in our system. 
Here $\varepsilon(q,\omega)$ stands for a dielectric function of the electron gas and 
$V_c({\bf q})=\frac{e^2}{2\bar{\varepsilon}_r\varepsilon_0 q}$ is the Fourier transform of the 
Coulomb potential between two electrons in 2D layer embedded between two dielectrics of 
average relative permitivity $\bar{\varepsilon}_r=(\varepsilon_{r1}+\varepsilon_{r2})/2$. 
We have assumed without loss of generality that the external field points in the $x$ direction (${\bf E}_0= {\bf \hat{x}} E_0$) and is parallel to the plane of our 2D electron gas.

In the case of randomly assembled impurities at positions ${\bf R}_j$, one can write for 
the scattering potential $V_i({\bf q})=-V_c({\bf q}) \sum_j e^{-i{\bf q}\cdot{\bf R}_j}$.
Note that we are assuming positively charged ($e$) impurities embedded in a see of negative electrons ($-e$). Then by averaging over random impurity positions one has 
$\langle |V_i({\bf q})|^2 \rangle = N_i \cdot V_c^2({\bf q})$, 
where $N_i$ is the number of impurities \cite{Mahan}. 

Equation (\ref{Absorp2D}) depends on the loss function 
$\Im\frac{1}{\varepsilon(q,\omega)}$ which generally contains contribution from single particle 
excitations and collective (plasmon) excitations. In this paper we focus solely on the 
plasmon contribution in which case one can write \cite{Pines}: 
\begin{equation}
\Im\frac{1}{\varepsilon(q,\omega)}=
\frac{-\pi}{ \frac{\partial\varepsilon}{\partial\omega}} 
\cdot \delta(\omega-\omega_q), 
\label{lossF_plasmon_omega}
\end{equation}
where $\omega_q$ is the plasmon frequency determined by the zero of the dielectric function: $\varepsilon(q,\omega_q)=0$. This term then represents the process where an incident photon excites plasmon of the same energy, through impurity scattering. 

The $\delta$-function from equation (\ref{lossF_plasmon_omega}) extracts only a single wavevector from the sum in equation (\ref{Absorp2D}), 
which corresponds to the plasmon wavevector at the given frequency $\omega$. 
Then one is left with integration over the angle $\varphi_{\bf q}$ which is straightforward 
to perform since $\int_0^{2\pi}d\varphi_{\bf q}\cdot q_x^2=\pi q^2$.

\begin{figure}
\centerline{
\mbox{\includegraphics[width=0.7\textwidth]{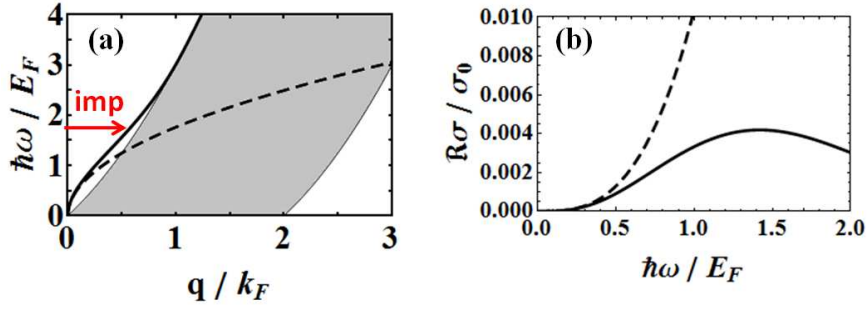}}
}
\caption{
Plasmon dispersion and optical conductivity for metallic monolayers.
In plot (a) we show plasmon dispersion relation within the Random Phase Approximation (solid line) and within the Drude model i.e in the small frequency limit (dashed line). Grey area denotes regime of single particle excitations. Random assembly of impurities break the translation invariance which allows a zero momentum photon to couple to a finite momentum plasmon (sketched by the red arrow). 
In plot (b) we show the optical absorption for the plasmon emission process through impurity scattering. We plot real part of the conductivity in units of $\sigma_0=\frac{e^2}{4\hbar}$, versus photon energy in units of Fermi energy $E_F$. One can see that the small frequency limit (dashed line) significantly overestimates the more exact RPA result (solid line).
}
\label{fig1}
\end{figure}

Finally, we plot the conductivity from expression (\ref{Absorp2D}) in Figure \ref{fig1} by using the dielectric function $\varepsilon(q,\omega)$ within Random Phase Approximation (RPA) given in Ref. \cite{Stern}. 
To represent the experiment \cite{Nagao2001}, which studied silver monolayer on a silicon substrate, we choose $\varepsilon_{r1}=\varepsilon_{Si}=12$, $\varepsilon_{r2}=1$, the effective mass $m^*=0.3 m$, where $m$ is the free electron mass, electron concentration $n=2\cdot 10^{13}$ cm$^{-2}$, and we assume the impurity concentration 
$n_i= 10^{12}$ cm$^{-2}$.

It is also convenient to look at the small frequency limit 
$(\hbar \omega\ll E_F)$ in which case only long wavelength $(q\ll q_F)$ plasmons contribute to the scattering. Here $E_F$ and $q_F$ stand for Fermi energy and Fermi momentum, respectively. In this limit, one can use a simple Drude model to obtain the dielectric function:
\begin{equation}
\varepsilon_D(q,\omega)=1-\frac{q}{\omega^2} \cdot
\frac{e^2n}{2\bar\varepsilon_r\varepsilon_0 m^*}.
\label{epsilon_Drude}
\end{equation}
In this case, plasmon dispersion is simply $\omega\propto\sqrt{q}$ and one can easily evaluate equations (\ref{Absorp2D}) and (\ref{lossF_plasmon_omega}) to obtain the conductivity:
\begin{equation}
\Re\sigma(\omega)=\frac{\pi e^2}{4\hbar} 
\frac{n_i}{q_{TF}^2}
\left( \frac{\hbar\omega}{E_F}\right)^3.
\label{Absorp2D_Drude}
\end{equation}
Here we have introduced the Thomas-Fermi wavevector: 
$q_{TF}=\frac{e^2 m^*}{2\pi\bar\varepsilon_r\varepsilon_0\hbar^2}$, while $n_i=N_i/\Omega$ stands for the impurity density.
From Fig. \ref{fig1} b we see that in the case of metallic monolayers the small frequency limit (dashed line) significantly 
overestimates the more exact RPA result (solid line).

\section{Graphene}
\label{sec:graphene}

\begin{figure}
\centerline{
\mbox{\includegraphics[width=0.7\textwidth]{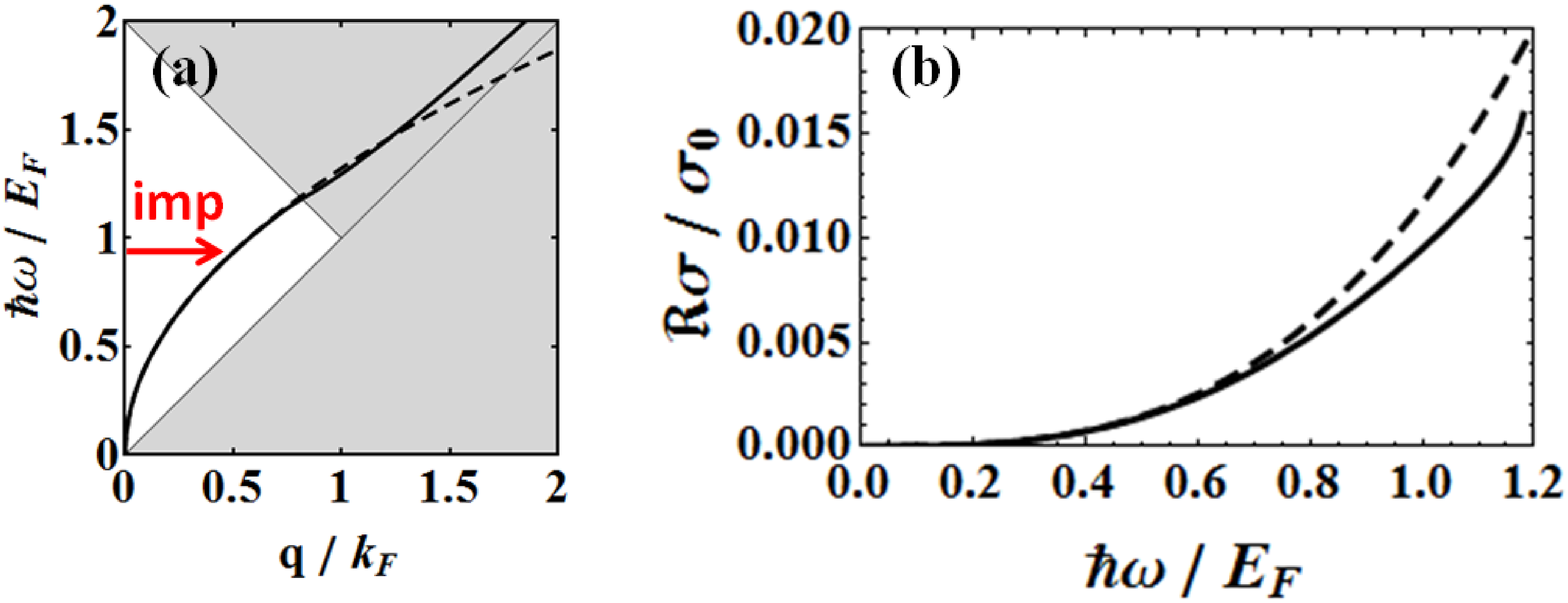}}
}
\caption{
Plasmon dispersion and optical conductivity for graphene sitting on the $SiO_2$ substrate with air above. 
In plot (a) we show plasmon dispersion within the Random Phase Approximation (solid line), and within the Drude model i.e. in the small frequency limit (dashed line). Grey area denotes regime of single particle excitations. Random assembly of impurities breaks the translation invariance, which allows a zero momentum photon to couple to a finite momentum plasmon (sketched by the red arrow). This process is possible only when the plasmon dispersion is outside of the grey area. Otherwise, plasmons are strongly damped due to single particle excitations (Landau damping). In plot (b) we show optical absorption for plasmon emission process through impurity scattering. We plot the real part of the conductivity in units of $\sigma_0=\frac{e^2}{4\hbar}$, versus photon energy in units of Fermi energy $E_F$.
One can see that the small frequency limit (dashed line) is very close to the more exact RPA result (solid line). This is related to the fact that in this case the plasmon dispersion from (a) is very well described by the small frequency limit. 
}
\label{fig2}
\end{figure}

Unfortunately, the trick that Hopfield used in the case of the parabolic dispersion  does not work for Dirac dispersion so one has to do the perturbation theory both in impurity scattering and in light scattering, while including the screening effect in every order of the perturbation theory. This is straightforward, but very tedious task, so we give the derivation of the optical absorption in the Appendix. Here we only write the final result: 
\begin{equation}
\Re\sigma(\omega)=-\frac{e^2v_F^2}{\omega}\frac{1}{\Omega}\sum_{\bf q} 
\frac{1}{\Omega} \left| \frac{V_i(q)}{\varepsilon(q)} \right|^2 F^2(q,\omega)
V_c(q) \cdot \Im\frac{1}{\varepsilon(q,\omega)},
\label{Absorp_Graphene}
\end{equation}
where we have assumed general impurity scattering Hamiltonian:
$H_i=\frac{1}{\Omega}\sum_{\bf q}V_i(q) e^{i{\bf q}\cdot{\bf r}}$ (see the Appendix for more details). In the case of charged impurities one has $\langle |V_i({\bf q})|^2 \rangle = N_i \cdot V_c^2({\bf q})$ after averaging over random impurity positions. Then, to find the contribution of plasmon emission process one can use equation (\ref{lossF_plasmon_omega}) and the dielectric function which is calculated in Ref. \cite{Hwang2007_plasmon} within the RPA. The resulting optical absorption, is plotted in Figure \ref{fig2}. To resemble parameters from the experiment \cite{Basov2008} we choose electron concentration $n=7\cdot 10^{12}$ cm$^{-2}$, and impurity concentration $n_i= 10^{12}$ cm$^{-2}$. Furthermore, we plot the case of graphene sitting on the $SiO_2$ substrate where $\bar\varepsilon_{r}=2.5$, but also the case of suspended graphene where $\bar\varepsilon_{r}=1$.

\begin{figure}
\centerline{
\mbox{\includegraphics[width=1\textwidth]{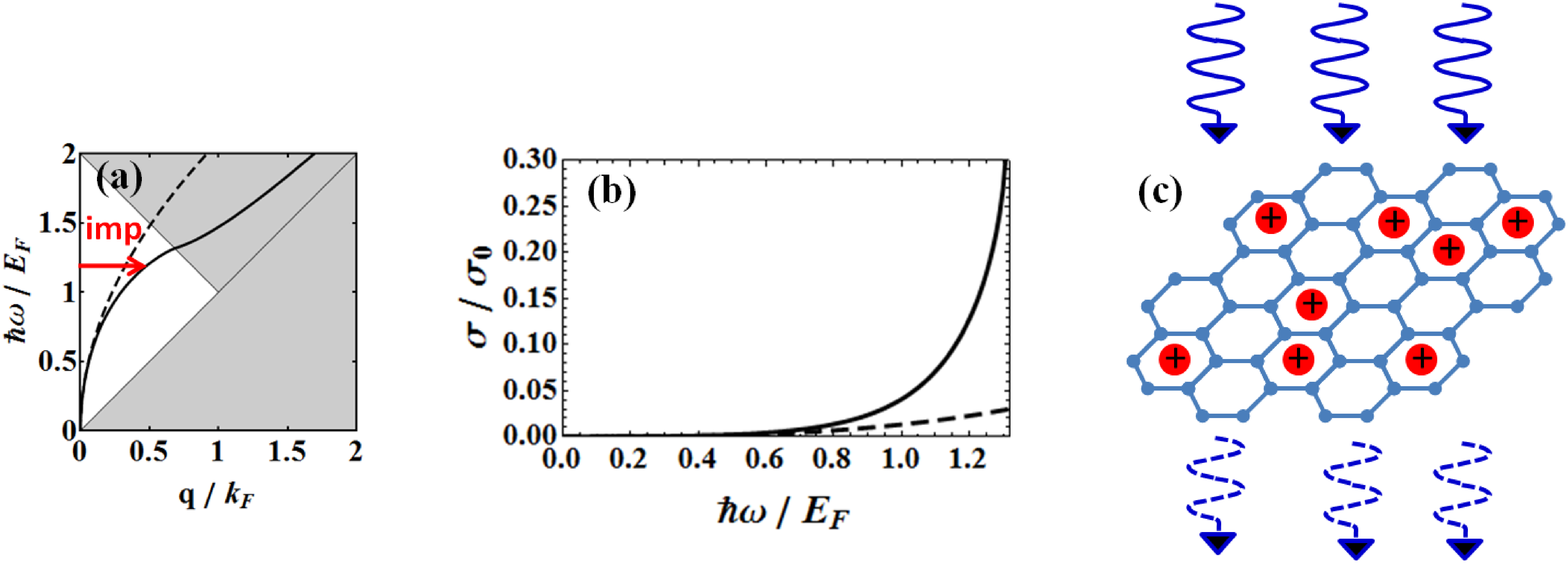}}
}
\caption{
Plasmon dispersion and optical conductivity for suspended graphene. 
In plot (a) we show plasmon dispersion within the Random Phase Approximation (solid line), and within the Drude model i.e. in the small frequency limit (dashed line). Grey area denotes regime of single particle excitations. Random assembly of impurities breaks the translation invariance, which allows a zero momentum photon to couple to a finite momentum plasmon (sketched by the red arrow). This process is possible only when the plasmon dispersion is outside of the grey area. Otherwise, plasmons are strongly damped due to single particle excitations (Landau damping). In plot (b) we show optical absorption for plasmon emission process through impurity scattering. We plot the real part of the conductivity in units of $\sigma_0=\frac{e^2}{4\hbar}$, versus photon energy in units of Fermi energy $E_F$.
One can see that the small frequency limit (dashed line) can be an order of magnitude lower that the more exact RPA result (solid line). 
The predicted loss mechanism should be observable in optical transmission measurements on suspended graphene, as sketched in plot (c). Red circules with crosses represent positively charged impurity ions that have donated electrons to the graphene plane. See text for details.
}
\label{fig3}
\end{figure}

It is also convenient to look at the small frequency limit 
$(\hbar \omega\ll E_F)$ in which case only long wavelength $(q\ll q_F)$ plasmons contribute to the scattering. Then, one can use a simple Drude model to obtain the dielectric function in graphene \cite{Jablan2009}:
\begin{equation}
\varepsilon_D(q,\omega)=1-\frac{q}{\omega^2} \cdot
\frac{e^2 v_F \sqrt{n}}{2\bar\varepsilon_r\varepsilon_0 \hbar\sqrt\pi}.
\label{epsilon_Drude_graphene}
\end{equation}
In this case the function $F$ takes a particularly simple expression (see the Appendix for more details):
$F(q,\omega)=\frac{-q_x}{\pi \hbar^2\omega v_F}$, and it is straightforward to evaluate expression (\ref{Absorp_Graphene}) to obtain:
\begin{equation}
\Re\sigma(\omega)=\frac{\pi e^2}{4\hbar} 
\frac{n_i}{q_{TF}^2}
\left( \frac{\hbar\omega}{E_F}\right)^3.
\label{Absorp2D_Drude_graphene}
\end{equation}
Note that this is the same result as in the case of metallic monolayers. This is expected because in the small frequency (long wavelength) limit, one does not expect to see specific details of the band structure. Of course, in the graphene case, the Thomas-Fermi wavevector is given by a different expression: 
$q_{TF}=\frac{e^2 q_F}{\pi\bar\varepsilon_r\varepsilon_0\hbar v_F}$. We would like to note that the small frequency limit in the case of graphene was also recently obtained by another group \cite{Sarma_small}. However, from Figure \ref{fig3} one can see that the small frequency limit can be very different from the more general RPA result. 

If we now compare our results [Figure \ref{fig2} (b)] with experiment \cite{Basov2008}, we see that this effect of plasmon emission is relatively small ($\Re\sigma<0.02\sigma_0$) compared to the experimental results ($\Re\sigma\approx 0.3\sigma_0$) in this regime. One might ask what are the other potentially strong scattering mechanisms? For example in experiment \cite{Basov2008}, graphene is sitting on $SiO_2$, which is a polar substrate, so there is a strong interaction of electrons with the surface polar phonons at energy $\hbar\omega_{SO}\approx 0.15$ eV. This is described by the Hamiltonian 
$H_{SO}=\frac{1}{\Omega}\sum_{\bf q}V_{SO}(q) 
\left( 
e^{i{\bf q}\cdot{\bf r}} a_{\bf q}^\dag + 
e^{-i{\bf q}\cdot{\bf r}} a_{\bf q}
\right)$,
where $a_{\bf q}^\dag$ is the phonon creation operator. For the square of the scattering potential we can write \cite{Massimo2001}: $V_{SO}^2(q)=\Omega\frac{e^2}{2\varepsilon_0}\hbar\omega_{SO}\left( \frac{1}{\varepsilon_r(\infty)+1}- \frac{1}{\varepsilon_r(0)+1}\right)
\frac{e^{-2qz}}{q}$. 
We use parameters from Ref. \cite{Massimo2001} for $SO$ scattering: $\varepsilon_r(0)=3.9$, $\varepsilon_r(\infty)=2.5$, and we assume that the Van der Waals distance between graphene and the substrate is $z=0.35$ nm. If we neglect the frequency dependence of $H_{SO}$, one can make an estimate of absorption simply by replacing $V_i(q)$ with $V_{SO}(q)$ in relation (\ref{Absorp_Graphene}). Strictly speaking this is valid only at large frequencies when $\omega\gg\omega_{SO}$, but it should give a reasonable estimate in the regime $\omega\approx 2\omega_{SO}$ which is the relevant regime in experiment \cite{Basov2008}. The resulting absorption is still extremely small ($\Re\sigma<0.003\sigma_0$) in the regime of interest ($\hbar\omega\approx E_F$).

Even though our analysis suggests that these loss mechanisms can not be distinguished from other loss mechanisms in current experiments involving graphene on a $SiO_2$ substrate, our calculations point out that they should be observable in suspended graphene (see Figure \ref{fig3}). Suspended graphene is a much cleaner system as one can eliminate all the scattering mechanisms that originate from the interaction with the substrate. Moreover, in optical transmission measurements on suspended graphene (sketched in Figure \ref{fig3} (c)) one does not need to consider optical absorption of the substrate. Suspended graphene can be doped by depositing electron-donor atoms like Sodium or Lithium. In that case one is left with impurity ions with the same number as the number of injected electrons. In Figure \ref{fig3} (b) we plot optical absorption in suspended graphene for identical impurity and electron concentrations $n_i=n=10^{12}$ cm$^{-2}$. One can see that there is a huge optical absorption through the plasmon emission channel as the real part of conductivity reaches $\Re\sigma\approx 0.3\sigma_0$. This would correspond to the $0.7\%$ reduction in the intensity of transmitted light, which could easily be observed as the $2.3\%$ reduction is already visible by naked eye \cite{Nair2008}. Finally we note that the small frequency limit (equation (\ref{Absorp2D_Drude_graphene})) underestimates the more exact RPA calculation (equation (\ref{Absorp_Graphene})) by an order of magnitude.

\section{Conclusion}
\label{sec:conclusion}

In conclusion, we have studied optical absorption of 2D electron gas in graphene and metallic monolayers with random distribution of charge impurities. This formalism can also treat other 2D electron systems like those found in heterostructures, single layer boron-nitride, or single layer molybdenum-disulphide where we expect similar behavior. 
Specifically, we have focused on a decay channel where an incident photon excites a plasmon through impurity scattering. For the graphene sitting on a $SiO_2$ substrate, we have also studied a decay channel where an incident photon excites a plasmon and an optical phonon of the polar substrate. The resulting optical absorption is more than one order of magnitude lower than the experimental results \cite{Basov2008}, and not enough to reconcile the difference between the theory \cite{Stauber2008,Carbotte2010,Peres2010,Vasko2012,Scharf2013} and the experiment \cite{Basov2008}. 
On the other hand we predict large optical absorption by plasmon emission via impurity scattering in suspended graphene. Thus we believe that these ideas can be tested in suspended graphene. Finally we note that for suspended graphene (metallic monolayers) the small frequency limit \cite{Sarma_small} gives an order of magnitude lower (larger) result than the more exact RPA calculation.

\appendix*
\section{Calculation of optical absorption in graphene}

We use single particle density matrix (SPDM) approach which is a convenient way to take into account both temperature and the Pauli principle. Equation of motion for SPDM $\rho$ is given by \cite{Ron1963}:
\begin{equation}
i\hbar\frac{\partial\rho}{\partial t}=\left[ H,\rho \right],
\label{eq_of_motion}
\end{equation}
where the Hamiltonian is given by 
\begin{equation}
H=H_0+H_l+H_i+H^s.
\label{Hamiltonian}
\end{equation}
Here $H_0$ represents kinetic energy of free electrons, $H_l$ describes scattering with light, $H_i$ scattering with impurities, and $H^s$ describes electron-electron interactions which we only take in the form of a self-consistent screening field. In the case of graphene, electrons are described by Dirac dispersion
\cite{Wallace1954,Semenoff1984}: 
\begin{equation}
H_0= \hbar v_F \mbox{\boldmath $\sigma$}\cdot{\bf k},
\label{Dirac}
\end{equation}
where $v_F=10^6$ m/s is Fermi velocity, $\bf k$ is electron wavevector, 
$\mbox{\boldmath $\sigma$}=\sigma_x{\bf\hat x}+\sigma_y{\bf\hat y}$, and $\sigma_{x,y}$ 
are the Pauli spin matrices. 
Let us denote by $| n {\bf k} \rangle$ eigenstates of $H_0$, where $n=1$ stands for the conduction band, and $n=-1$ for the valence band. Then the eigenvalues of $H_0$ are given by Dirac cones: $E_{n{\bf k}}=n \hbar v_F |{\bf k}|$. 
If we now introduce a light source described by the electric field 
${\bf E}(t)= {\bf \hat x} E_0 e^{-i\omega t} + c.c.$, 
then scattering with light is determined by the Hamiltonian:
\begin{equation}
H_l=-i\frac{e v_F}{\omega} \sigma_x E_0 e^{-i\omega t}+c.c., 
\label{light_scattering}
\end{equation} 
where $-e$ is the electron charge. 
Furthermore, we can write the Hamiltonian for impurity scattering as a Fourier sum over wavevectors 
$\bf q$: 
\begin{equation}
H_i=\frac{1}{\Omega}\sum_{\bf q} V_i(q)e^{i{\bf q}\cdot{\bf r}}, 
\label{impurity_scattering}
\end{equation}
where $\Omega$ is total area of our graphene flake, $\bf r$ is the position operator, and $V_i(q)$ is the Fourier transform of the scattering potential. Here we assume a general scattering potential and only later we will specify $V_i(q)$ for the case of charged impurity scattering and surface polar phonon scattering. 
Finally, one can also write the screening field as a Fourier sum: 
\begin{equation}
H^s=\frac{1}{\Omega}\sum_{\bf q} V^s({\bf q})e^{i{\bf q}\cdot{\bf r}}e^{-i\omega t}+c.c.,
\label{screening_field}
\end{equation}
but one has to keep in mind that different orders of the perturbation expansion will have different time dependence (frequencies). 
Here, the screening field is taken as a self-consistent electrostatic field that the electrons induce on themselves, so one can write $V^s({\bf q})=V_c(q)n({\bf q})$, where 
$n({\bf q})$ is the Fourier transform of the electron density, and $V_c(q)$ is the Fourier transform of the Coulomb potential between two electrons. For a 2D electron gas embedded between two dielectrics of relative permitivity $\bar{\varepsilon}_r=(\varepsilon_{r1}+\varepsilon_{r2})/2$, one can write: 
$V_c(q)=\frac{e^2}{2\bar{\varepsilon}_r\varepsilon_0 q}$. Note that this is valid only in the electrostatic limit $q\gg\omega/c$ which is the relevant regime for our case. Furthermore, since 
$n({\bf q})=Tr\left\{e^{-i{\bf q}\cdot{\bf r}}\rho\right\}$, one can write for the screening field:
\begin{equation}
V^s({\bf q})=V_c(q)\cdot 4\sum_{n_1 n_2 {\bf k}} 
\langle n_1 {\bf k}| e^{-i{\bf q}\cdot{\bf r}} | n_2 {\bf k+q} \rangle 
\langle n_2 {\bf k+q} | \rho | n_1 {\bf k} \rangle ,
\label{screening_field_tr}
\end{equation}
where we have taken into account 2 spin and 2 valley degeneracies.
We are now interested in calculating the current response up to the linear order in the external electric field ${\bf E}(t)$. Since the electric field is uniform in the graphene plane, we are only interested in the ${\bf q}=0$ term, and the current density operator is given by 
${\bf j}_{op}=-\frac{e v_F }{\Omega}\mbox{\boldmath $\sigma$}$.
The induced current will have only the $x$ component, since the 
electric field points in the $x$ direction.
Finally, the induced current density is given by
${\bf j}=Tr\left\{{\bf j}_{op}\;\rho\right\}$, so we can write:
\begin{equation}
j_x=-\frac{e v_F }{\Omega} \cdot 4 \sum_{n_1 n_2 {\bf k}} 
\langle n_1 {\bf k}| \sigma_x | n_2 {\bf k} \rangle
\langle n_2 {\bf k} | \rho | n_1 {\bf k} \rangle.  
\label{current_tr}
\end{equation}
To include also impurity scattering, we need to calculate the induced current up to the second order in $V_i({\bf q})$. In other words we need to do a perturbation expansion of SPDM:
\begin{eqnarray}
\rho=\rho_0+\rho_l+\rho_i+\rho_{li}+\rho_{lii},
\label{SPDM_expand}
\end{eqnarray}
where $\rho_0$ is the equilibrium solution to equation (\ref{eq_of_motion}) for independent Dirac electrons in the absence of impurity scattering and light scattering, $\rho_l\propto H_l$ is solution of equation (\ref{eq_of_motion}) correct up to a linear order in light scattering, $\rho_i\propto H_i$ is solution up to the linear order in impurity scattering, $\rho_{li}\propto H_l\cdot H_i$ is solution up to linear order both in light scattering and impurity scattering, and $\rho_{lii}\propto H_l\cdot H_i^2$ is solution up to the linear order in light scattering and quadratic in impurity scattering. Using equation (\ref{eq_of_motion}), we can now write the equation of motion for every order of SPDM expansion:
\begin{equation}
i\hbar\frac{\partial\rho_0}{\partial t}=\left[ H_0,\rho_0 \right],
\label{eq_of_motion_0}
\end{equation}
\begin{equation}
i\hbar\frac{\partial\rho_l}{\partial t}=\left[ H_0,\rho_l \right]+
\left[ H_l+H_l^s,\rho_0 \right],
\label{eq_of_motion_l}
\end{equation}
\begin{equation}
i\hbar\frac{\partial\rho_i}{\partial t}=\left[ H_0,\rho_i \right]+\left[ H_i+H_i^s,\rho_0 \right],
\label{eq_of_motion_i}
\end{equation}
\begin{equation}
i\hbar\frac{\partial\rho_{li}}{\partial t}=
\left[ H_0,\rho_{li} \right]+
\left[ H_i+H_i^s,\rho_l \right]+
\left[ H_l+H_l^s,\rho_i \right]+
\left[ H_{li}^s,\rho_0 \right], \; \mbox{and}
\label{eq_of_motion_li}
\end{equation}
\begin{equation}
i\hbar\frac{\partial\rho_{lii}}{\partial t}=
\left[ H_0,\rho_{lii} \right]+
\left[ H_i+H_i^s,\rho_{li} \right]+
\left[ H_{li}^s,\rho_i \right]+
\left[ H_{lii}^s,\rho_0 \right].
\label{eq_of_motion_lii}
\end{equation}
The equilibrium solution of equation (\ref{eq_of_motion_0}) describes the free electrons and is given by:
\begin{equation}
\langle n_2 {\bf k+q} |\rho_0| n_1 {\bf k} \rangle = 
\delta_{n_1,n_2}\delta_{{\bf q},0} \cdot f_{n_1{\bf k}}, 
\label{rho_0}
\end{equation}
where $\delta_{a,b}$ is the Kronecker delta symbol and 
$f_{n{\bf k}}=\left[ e^{(E_{n{\bf k}}-E_F)/kT}+1 \right]^{-1}$ is the Fermi-Dirac distribution at temperature $T$ and Fermi energy $E_F$. Using relation (\ref{rho_0}) we can write the solution of equation (\ref{eq_of_motion_l}) as:
\begin{equation}
\langle n_2 {\bf k+q} |\rho_l| n_1 {\bf k} \rangle =
-i\frac{e v_F}{\omega} E_0 \cdot \delta_{{\bf q},0}
\langle n_2 {\bf k} |\sigma_x| n_1 {\bf k} \rangle 
\frac{f_{n_1{\bf k}}-f_{n_2{\bf k}}}{\hbar\omega+E_{n_1{\bf k}}-E_{n_2{\bf k}}},
\label{rho_l}
\end{equation}
which is a stady-state solution of SPDM that oscillates at frequency $\omega$.
Here we have used the following relation: 
$\langle n_2 {\bf k+q} |\sigma_x| n_1 {\bf k} \rangle=\delta_{{\bf q},0}
\langle n_2 {\bf k} |\sigma_x| n_1 {\bf k} \rangle$. We have neglected the screening field $H_l^s$ in equation (\ref{eq_of_motion_l}) since the 2D electron gas can not screen the uniform electric field. This can be seen below from equation (\ref{epsilon_approx}) which gives the dielectric function of graphene in the long wavelenght limit. One can immediately see that $\varepsilon(q=0,\omega)=1$ which means that there is no screening in the $q=0$ limit.

Let us now focus on equation (\ref{eq_of_motion_i}). We can introduce a self-consistent Hamiltonian $H_i^{sc}=H_i+H_i^s$, and write:  
$H_i^{sc}=\frac{1}{\Omega}\sum_{\bf q} V_i^{sc}(q)e^{i{\bf q}\cdot{\bf r}}$, where 
$V_i^{sc}=V_i+V_i^s$ is a self-consistent scattering potential 
that consists of a bare impurity scattering potential $V_i$, and a screening field 
$V_i^s$. By solving equations (\ref{screening_field_tr}) and (\ref{eq_of_motion_i}) in a self-consistent way one can show that $V_i^{sc}(q)=V_i(q)/\varepsilon(q)$, where $\varepsilon(q)$ is the static dielectric function. 
The dynamic dielectric function of graphene is generally
\begin{equation}
\varepsilon(q,\omega)=1-V_c(q)\frac{4}{\Omega}\sum_{n_1n_2{\bf k}}
\frac{f_{n_1{\bf k}}-f_{n_2{\bf k+q}}}
{\hbar\omega+E_{n_1{\bf k}}-E_{n_2{\bf k+q}}}
|\langle n_2 {\bf k+q} |e^{i{\bf q}\cdot{\bf r}}| n_1 {\bf k} \rangle|^2,
\label{dielectric_function}
\end{equation} 
and one can simply check that $\varepsilon(q)=\varepsilon(q,\omega=0)$. Finally, the solution to equation (\ref{eq_of_motion_i}) can be written as
\begin{equation}
\langle n_2 {\bf k+q} |\rho_i| n_1 {\bf k} \rangle =
\frac{1}{\Omega}\frac{V_i(q)}{\varepsilon(q)}
\frac{f_{n_1{\bf k}}-f_{n_2{\bf k+q}}}{E_{n_1{\bf k}}-E_{n_2{\bf k+q}}}
\langle n_2 {\bf k+q} |e^{i{\bf q}\cdot{\bf r}}| n_1 {\bf k} \rangle.
\label{rho_i}
\end{equation}
To solve the next order of perturbation theory $\rho_{li}$ we need to include the screening field described by a Hamiltonian 
$H_{li}^s=\frac{1}{\Omega}\sum_{\bf q} V_{li}^s({\bf q})e^{i{\bf q}\cdot{\bf r}}e^{-i\omega t}+c.c.$. One can then solve equation (\ref{eq_of_motion_li}) by using results (\ref{rho_0}), (\ref{rho_l}) and (\ref{rho_i}) to obtain:
\begin{align}
\langle n_2 {\bf k+q} |\rho_{li}| n_1 {\bf k} \rangle & =
\frac{1}{\Omega}V_{li}^s({\bf q})
\frac{f_{n_1{\bf k}}-f_{n_2{\bf k+q}}}
{\hbar\omega+E_{n_1{\bf k}}-E_{n_2{\bf k+q}}}
\langle n_2 {\bf k+q} |e^{i{\bf q}\cdot{\bf r}}| n_1 {\bf k} \rangle
\nonumber \\
& +\frac{1}{\Omega}\frac{V_i(q)}{\varepsilon(q)} (-i)\frac{e v_F}{\omega} E_0 
\frac{1}{\hbar\omega+E_{n_1{\bf k}}-E_{n_2{\bf k+q}}} \times
\nonumber \\
&  \times \left(
\sum_{n_3} \langle n_2 {\bf k+q} |e^{i{\bf q}\cdot{\bf r}}| n_3 {\bf k} \rangle
\langle n_3 {\bf k} |\sigma_x| n_1 {\bf k} \rangle 
\frac{f_{n_1{\bf k}}-f_{n_3{\bf k}}}
{\hbar\omega+E_{n_1{\bf k}}-E_{n_3{\bf k}}} \right.
\nonumber \\
& \;\;\; \left. - \sum_{n_3} \langle n_2 {\bf k+q} |\sigma_x| n_3 {\bf k+q} \rangle
\langle n_3 {\bf k+q} |e^{i{\bf q}\cdot{\bf r}}| n_1 {\bf k} \rangle 
\frac{f_{n_3{\bf k+q}}-f_{n_2{\bf k+q}}}
{\hbar\omega+E_{n_3{\bf k+q}}-E_{n_2{\bf k+q}}} \right.
\nonumber \\
& \;\;\; \left. + \sum_{n_3} \langle n_2 {\bf k+q} |\sigma_x| n_3 {\bf k+q} \rangle
\langle n_3 {\bf k+q} |e^{i{\bf q}\cdot{\bf r}}| n_1 {\bf k} \rangle 
\frac{f_{n_1{\bf k}}-f_{n_3{\bf k+q}}}
{E_{n_1{\bf k}}-E_{n_3{\bf k+q}}} \right.
\nonumber \\
&  \;\;\; \left. - \sum_{n_3} \langle n_2 {\bf k+q} |e^{i{\bf q}\cdot{\bf r}}| n_3 {\bf k} \rangle
\langle n_3 {\bf k} |\sigma_x| n_1 {\bf k} \rangle 
\frac{f_{n_3{\bf k}}-f_{n_2{\bf k+q}}}
{E_{n_3{\bf k}}-E_{n_2{\bf k+q}}}
\right) .
\label{rho_li}
\end{align}
Next, one can use relation (\ref{screening_field_tr}) to obtain the screening field in a self-consistent way:
\begin{align}
V_{li}^s({\bf q})&= 
\frac{V_i(q)}{\varepsilon(q)}\frac{V_c(q)}{\varepsilon(q,\omega)}
(-i)\frac{e v_F}{\omega} E_0 
\frac{4}{\Omega}\sum_{n_1n_2n_3{\bf k}}
\frac{\langle n_1 {\bf k} |e^{-i{\bf q}\cdot{\bf r}}| n_2 {\bf k+q} \rangle}
{\hbar\omega+E_{n_1{\bf k}}-E_{n_2{\bf k+q}}} \times
\nonumber \\
&  \times \left(
\langle n_2 {\bf k+q} |\sigma_x| n_3 {\bf k+q} \rangle
\langle n_3 {\bf k+q} |e^{i{\bf q}\cdot{\bf r}}| n_1 {\bf k} \rangle 
\frac{f_{n_1{\bf k}}-f_{n_3{\bf k+q}}}
{E_{n_1{\bf k}}-E_{n_3{\bf k+q}}} \right.
\nonumber \\
&  \;\;\; \left. 
-\langle n_2 {\bf k+q} |e^{i{\bf q}\cdot{\bf r}}| n_3 {\bf k} \rangle
\langle n_3 {\bf k} |\sigma_x| n_1 {\bf k} \rangle 
\frac{f_{n_3{\bf k}}-f_{n_2{\bf k+q}}}
{E_{n_3{\bf k}}-E_{n_2{\bf k+q}}}
\right) ,
\label{screening_field_li}
\end{align}
where $\varepsilon(q,\omega)$ is the dynamic dielectric function given in (\ref{dielectric_function}). Note that the terms containing 
 $f_{n_1{\bf k}}-f_{n_3{\bf k}}$ and
$f_{n_3{\bf k+q}}-f_{n_2{\bf k+q}}$ have disappeared after summation over $n_1,n_2,n_3$ and $\bf k$. One can also demonstrate the following important property: $V_{li}^s(-{\bf q})=-V_{li}^s({\bf q})$. Finally, one can use equation (\ref{eq_of_motion_lii}) to find $\rho_{lii}$, and equation (\ref{current_tr}) to find the induced current up to the first order in light scattering, and the second order in impurity scattering:
\begin{align}
j_x^{lii} & = -\frac{e v_F}{\Omega}\cdot 4\sum_{n_1n_2n_3{\bf k},{\bf q}}
\frac{\langle n_1{\bf k}|\sigma_x|n_2 {\bf k} \rangle}
{\hbar\omega+E_{n_1{\bf k}}-E_{n_2{\bf k}}} \times
\nonumber \\
& \times \left( \frac{1}{\Omega}\frac{V_i(q)}{\varepsilon(q)}
\langle n_2 {\bf k} |e^{-i{\bf q}\cdot{\bf r}}| n_3 {\bf k+q} \rangle
\langle n_3 {\bf k+q} |\rho_{li}| n_1 {\bf k} \rangle 
\right.
\nonumber \\
& \;\;\; - \frac{1}{\Omega}\frac{V_i(q)}{\varepsilon(q)}
\langle n_2 {\bf k} |\rho_{li}| n_3 {\bf k-q} \rangle
\langle n_3 {\bf k-q} |e^{-i{\bf q}\cdot{\bf r}}| n_1 {\bf k} \rangle 
\nonumber \\
& \;\;\; + \frac{1}{\Omega}V_{li}^s(-{\bf q})
\langle n_2 {\bf k} |e^{-i{\bf q}\cdot{\bf r}}| n_3 {\bf k+q} \rangle
\langle n_3 {\bf k+q} |\rho_i| n_1 {\bf k} \rangle 
\nonumber \\
& \left. \;\;\; - \frac{1}{\Omega}V_{li}^s(-{\bf q})
\langle n_2 {\bf k} |\rho_i| n_3 {\bf k-q} \rangle
\langle n_3 {\bf k-q} |e^{-i{\bf q}\cdot{\bf r}}| n_1 {\bf k} \rangle \right).
\label{current_lii}
\end{align}
Note that we have neglected the screening field $H_{lii}^s$ since we need only the $q=0$ component of $\rho_{lii}$ to obtain $j_x^{lii}$, and there is no screening in the 2D electron gas in the $q=0$ case.
Also note that we have skipped the lower orders in the induced current since one can generally show that $j_x^{li}=0$. On the other hand $j_x^l\neq 0$ but we are here interested in the optical absorption below interband threshold $\hbar\omega<2E_F$, where $\Re j_x^l=0$. Finally, to evaluate the current component $j_x^{lii}$ from expression (\ref{current_lii}) we need to use expression (\ref{rho_i}) for $\rho_i$ and expression (\ref{rho_li}) for $\rho_{li}$. The resulting conductivity is:
\begin{align}
& \sigma_{lii}(\omega)=i\frac{e^2v_F^2}{\omega}\frac{1}{\Omega}\sum_{\bf q}
\frac{1}{\Omega} \left|\frac{V_i(q)}{\varepsilon(q)}\right|^2 \times
\nonumber \\
& \times \left( \frac{V_c(q)}{\varepsilon(q,\omega)}F^2({\bf q},\omega)
+\frac{4}{\Omega}\sum_{n_1n_2n_3n_4{\bf k}} 
G(n_1,n_2,n_3,{\bf k},{\bf q},\omega) \cdot 
H(n_1,n_2,n_4,{\bf k},{\bf q},\omega)
\right),
\label{conductivity_lii}
\end{align}
where the functions $F$, $G$ and $H$ are given by the following expressions:
\begin{align}
F({\bf q},\omega) & = -\frac{4}{\Omega}\sum_{n_1n_2n_3{\bf k}}
\frac{f_{n_1{\bf k}}-f_{n_2{\bf k+q}}}{E_{n_1{\bf k}}-E_{n_2{\bf k+q}}} \times
\nonumber \\
& \times \left( 
\frac{\langle n_3 {\bf k} |e^{-i{\bf q}\cdot{\bf r}}| n_2 {\bf k+q} \rangle}
{\hbar\omega+E_{n_3{\bf k}}-E_{n_2{\bf k+q}}}
\langle n_2 {\bf k+q} |e^{i{\bf q}\cdot{\bf r}}| n_1 {\bf k} \rangle
\langle n_1 {\bf k} |\sigma_x| n_3 {\bf k} \rangle \right.
\nonumber \\
& \left. \;\;\; - \langle n_1 {\bf k} |e^{-i{\bf q}\cdot{\bf r}}| n_2 {\bf k+q} \rangle
\frac{\langle n_2 {\bf k+q} |e^{i{\bf q}\cdot{\bf r}}| n_3 {\bf k} \rangle}
{-\hbar\omega+E_{n_3{\bf k}}-E_{n_2{\bf k+q}}}
\langle n_3 {\bf k} |\sigma_x| n_1 {\bf k} \rangle
\right).
\label{F}
\end{align}
\begin{align}
G(n_1,n_2,n_3,{\bf k},{\bf q},\omega) & =
\frac{1}{\hbar\omega+E_{n_1{\bf k}}-E_{n_2{\bf k+q}}} \times
\nonumber \\
& \times \left( 
\langle n_3 {\bf k} |e^{-i{\bf q}\cdot{\bf r}}| n_2 {\bf k+q} \rangle
\frac{\langle n_1 {\bf k} |\sigma_x| n_3 {\bf k} \rangle}
{\hbar\omega+E_{n_1{\bf k}}-E_{n_3{\bf k}}} 
\right.
\nonumber \\
& \left. \;\;\; - \langle n_1 {\bf k} |e^{-i{\bf q}\cdot{\bf r}}| n_3 {\bf k+q} \rangle
\frac{\langle n_3 {\bf k+q} |\sigma_x| n_2 {\bf k+q} \rangle}
{\hbar\omega+E_{n_3{\bf k+q}}-E_{n_2{\bf k+q}}} 
\right).
\label{G}
\end{align}
\begin{align}
H(n_1,n_2,n_4,{\bf k},{\bf q},\omega) & =
\langle n_2 {\bf k+q} |e^{i{\bf q}\cdot{\bf r}}| n_4 {\bf k} \rangle
\langle n_4 {\bf k} |\sigma_x| n_1 {\bf k} \rangle \times
\nonumber \\ 
& \times \left( 
\frac{f_{n_1{\bf k}}-f_{n_4{\bf k}}}
{\hbar\omega+E_{n_1{\bf k}}-E_{n_4{\bf k}}} 
-\frac{f_{n_4{\bf k}}-f_{n_2{\bf k+q}}}
{E_{n_4{\bf k}}-E_{n_2{\bf k+q}}} 
\right) 
\nonumber \\ 
& + \langle n_2 {\bf k+q} |\sigma_x| n_4 {\bf k+q} \rangle
\langle n_4 {\bf k+q} |e^{i{\bf q}\cdot{\bf r}}| n_1 {\bf k} \rangle \times
\nonumber \\ 
& \times \left( 
-\frac{f_{n_4{\bf k+q}}-f_{n_2{\bf k+q}}}
{\hbar\omega+E_{n_4{\bf k+q}}-E_{n_2{\bf k+q}}} 
+\frac{f_{n_1{\bf k}}-f_{n_4{\bf k+q}}}
{E_{n_1{\bf k}}-E_{n_4{\bf k+q}}} 
\right).
\label{H}
\end{align}
However, if we are interested only in the contribution from the collective excitations, we can neglect the single particle excitations to obtain:
\begin{equation}
\Re\sigma(\omega)=-\frac{e^2v_F^2}{\omega}\frac{1}{\Omega}\sum_{\bf q} 
\frac{1}{\Omega} \left| \frac{V_i(q)}{\varepsilon(q)} \right|^2 
F^2({\bf q},\omega) V_c(q) \cdot \Im\frac{1}{\varepsilon(q,\omega)}.
\label{absorb_plasmon_contrib}
\end{equation}
Note that this is the complete expression for the real part of conductivity, i.e. $\Re\sigma(\omega)=\Re\sigma_{lii}(\omega)$ since $\Re\sigma_l(\omega)=0$ in this regime, and generally $\Re\sigma_{li}(\omega)=0$. Then, since we are only interested in the plasmon contribution one can write the loss function as
\begin{equation}
\Im\frac{1}{\varepsilon(q,\omega)}=
\frac{-\pi}{ \frac{\partial\varepsilon}{\partial\omega}} 
\cdot \delta(\omega-\omega_q)=
\frac{\pi}{ \frac{\partial\varepsilon}{\partial q}} 
\cdot \delta(q-q_\omega), 
\label{loss_function_plasmon}
\end{equation}
where $\omega_q$ is plasmon frequency at a given wavevector $q$, and
$q_\omega$ is plasmon wavevector at a given frequency $\omega$, which is determined by the zero of the dielectric function: $\varepsilon(q,\omega_q)=\varepsilon(q_\omega,\omega)=0$. 
Note that $\delta$-function from equation (\ref{loss_function_plasmon}) extracts only a single wavevector from the integral $\int dq$ in equation (\ref{absorb_plasmon_contrib}). Moreover, one can explicitly perform the remaining integral $\int d\varphi_{\bf q}$. To demonstrate this we start by writing the expression for the Dirac wave function in coordinate representation:
\begin{equation}
\psi_{n,\bf k}({\bf r})= 
\langle {\bf r} | n{\bf k} \rangle=
\frac{1}{\sqrt{2\Omega}}
\left( {\begin{array}{c}
n \\
e^{i \varphi_{\bf k}} \\
\end{array}} \right)
e^{i {\bf k}\cdot{\bf r}}.
\label{wavef-Dirac}
\end{equation}
It is straightforward to calculate the following matrix elements:
\begin{equation}
\langle n {\bf k} | e^{-i{\bf q}\cdot{\bf r}} | n' {\bf k+q} \rangle =
\frac{1}{2}\left(  n n' + e^{-i\varphi_{\bf k}+i\varphi_{\bf k+q}}  \right),
\label{matrix1}
\end{equation}
\begin{equation}
\langle  n' {\bf k+q}| e^{i{\bf q}\cdot{\bf r}} |  n {\bf k} \rangle =
\frac{1}{2}\left(  n n' + e^{i\varphi_{\bf k}-i\varphi_{\bf k+q}}  \right),
\label{matrix2}
\end{equation}
\begin{equation}
\langle  n' {\bf k}| \sigma_x |  n {\bf k} \rangle =
\frac{1}{2}\left(  n e^{-i\varphi_{\bf k}} + n' e^{i\varphi_{\bf k}} \right).
\label{matrix3}
\end{equation}
Furthermore, the product of the last three terms can be written as:
\begin{align}
& \langle n {\bf k} | e^{-i{\bf q}\cdot{\bf r}} | n' {\bf k+q} \rangle
\langle  n' {\bf k+q}| e^{i{\bf q}\cdot{\bf r}} |  n'' {\bf k} \rangle 
\langle  n'' {\bf k}| \sigma_x |  n {\bf k} \rangle =
\nonumber \\
& = \left( \frac{1}{4}(1+n n'') + 
\frac{n'}{4}(n+n'')\frac{k+q\cos\varphi}{|{\bf k+q}|} +
i\frac{n'}{4}(n-n'')\frac{q\sin\varphi}{|{\bf k+q}|} \right) 
\nonumber \\
& \times \left(
\frac{\cos\varphi}{2} 
\left[ n'' e^{i\varphi_{\bf q}} + n e^{-i\varphi_{\bf q}} \right] +
i \frac{\sin\varphi}{2} 
\left[ n'' e^{i\varphi_{\bf q}} - n e^{-i\varphi_{\bf q}} \right]
\right),
\label{product_matrix}
\end{align}
where $\varphi=\varphi_{\bf k}-\varphi_{\bf q}$. Finally one can show that:
\begin{equation}
F({\bf q},\omega)=\tilde{F}(q,\omega)\cdot\cos\varphi_{\bf q}
\label{skalping_F}
\end{equation}
where $\tilde{F}(q,\omega)$ depends only on the magnitude of the wavevector $q$ and is given by the following expression:
\begin{align}
\tilde{F}(q,\omega)=
 - & \frac{4}{\Omega}\sum_{n_1n_2n_3{\bf k}} 
\frac{f_{n_1{\bf k}}-f_{n_2{\bf k+q}}}{E_{n_1{\bf k}}-E_{n_2{\bf k+q}}}
\left( \frac{1}{\hbar\omega+E_{n_3{\bf k}}-E_{n_2{\bf k+q}}}-
\frac{1}{-\hbar\omega+E_{n_3{\bf k}}-E_{n_2{\bf k+q}}} \right)
\nonumber \\
& \times \left\{
n_1(n_1+n_3)\left( n_1 + n_2\frac{k+q\cos\varphi}{|{\bf k+q}|} \right) \frac{\cos\varphi}{4} +
n_1(n_1-n_3)n_2\frac{q\sin\varphi}{|{\bf k+q}|}
\frac{\sin\varphi}{4}
\right\}.
\label{F_skalped}
\end{align}
Now one can indeed see that the that integration over $d\varphi_{\bf q}$ in equation (\ref{absorb_plasmon_contrib}) simply contributes with the following factor: 
$\int_0^{2\pi}d\varphi_{\bf q}\cos^2\varphi_{\bf q}=\pi$.
Finally, equation (\ref{absorb_plasmon_contrib}) is reduced to the following expression:
\begin{equation}
\Re\sigma(\omega)=\left. -\frac{e^2v_F^2}{\omega}\frac{1}{4\pi} q
\frac{1}{\Omega} \left| \frac{V_i(q)}{\varepsilon(q)} \right|^2 
\tilde{F}^2(q,\omega) V_c(q) \cdot
\frac{\pi}{ \frac{\partial\varepsilon(q,\omega)}{\partial q}} \right|_{pl},
\label{absorb_plasmon_contrib_reduced}
\end{equation}
where $q$ is the plasmon wavevector at the frequency $\omega$. To evaluate this expression one  needs to calculate the double integral 
$\int dk \int d\varphi_{\bf k} $ to evaluate the function $\tilde{F}(q,\omega)$. This can be further simplified at zero temperature when the Fermi-Dirac distrubution is a step function. 
In that case, we can group $(n_1,n_2,n_3)$ and $(-n_1,-n_2,-n_3)$ terms in equation (\ref{F_skalped}) to obtain:
\begin{align}
\tilde{F}(q,\omega)=
 - & \frac{2}{\Omega}\sum_{n_1n_2n_3{\bf k}} 
\frac{f_{\bf k}-f_{\bf k+q}}{E_{n_1{\bf k}}-E_{n_2{\bf k+q}}}
\left( \frac{1}{\hbar\omega+E_{n_3{\bf k}}-E_{n_2{\bf k+q}}}-
\frac{1}{-\hbar\omega+E_{n_3{\bf k}}-E_{n_2{\bf k+q}}} \right)
\nonumber \\
& \times \left\{
n_1(n_1+n_3)\left( n_1 + n_2\frac{k+q\cos\varphi}{|{\bf k+q}|} \right) \frac{\cos\varphi}{4} +
n_1(n_1-n_3)n_2\frac{q\sin\varphi}{|{\bf k+q}|}
\frac{\sin\varphi}{4}
\right\},
\label{F_skalped_simple}
\end{align}
where $f_{\bf k}=f_{n=1,{\bf k}}$ stands for the Fermi-Dirac distribution of the conduction band, and we have assumed electron doping i.e. $E_F>0$. We perform a numerical integration to evaluate the function $\tilde{F}(q,\omega)$; however, one can obtain a closed expression in the small frequency limit when 
$\hbar\omega\ll E_F$. In that case, only intraband transitions contribute and one can set $n_1=n_2=n_3=1$ in equation (\ref{F_skalped_simple}). Furthermore, in that case plasmon wavevector $q$ is much smaller than the Fermi wavevector $q_F$ so one can use the long wavelength expansions:
\begin{equation}
E_{\bf k}-E_{\bf k+q}=-\nabla_{\bf k}E_{\bf k}\cdot{\bf q},\; \mbox{and} 
\label{energy_expand}
\end{equation}
\begin{equation}
f_{\bf k}-f_{\bf k+q}=-\frac{\partial f}{\partial E}
(\nabla_{\bf k}E_{\bf k}\cdot{\bf q}). 
\label{f_expand}
\end{equation}
Next, it is straightforward to perform integration in equation (\ref{F_skalped_simple}) to obtain long wavelength (small frequency) approximation:
\begin{equation}
\tilde{F}(q,\omega)=-\frac{q}{\pi\hbar^2\omega v_F}. 
\label{F_skalped_approx}
\end{equation}
In a similar manner, from equation (\ref{dielectric_function}), one can obtain the dielectric function in this approximation:
\begin{equation}
\varepsilon(q,\omega)=1-\frac{q}{\omega^2} \cdot
\frac{e^2 v_F \sqrt{n}}{2\bar\varepsilon_r\varepsilon_0 \hbar\sqrt\pi},
\label{epsilon_approx}
\end{equation}
which is just the Drude model for the dielectric function in graphene \cite{Jablan2009}. 

Finally, from equation (\ref{absorb_plasmon_contrib_reduced}) we obtain optical absorption in the small frequency limit:
\begin{equation}
\Re\sigma(\omega)=\frac{\pi e^2}{4\hbar} 
\frac{n_i}{q_{TF}^2}
\left( \frac{\hbar\omega}{E_F}\right)^3,
\label{absorb_plasmon_contrib_reduced_small}
\end{equation}
where $q_{TF}=\frac{e^2 q_F}{\pi\bar\varepsilon_r\varepsilon_0\hbar v_F}$ 
is the Thomas-Fermi wavevector.


\acknowledgments
This work was supported in part by the Unity through Knowledge Fund (UKF grant No 5/13).
The work of M. Solja\v{c}i\'{c} was supported in part by the MIT S3TEC Energy Research Frontier Center of the Department of Energy under Grant DE-SC0001299. This work was also supported in part by the Army Research Office through the Institute for Soldier Nanotechnologies under Contract W911NF-13-D-0001.


\end{document}